\documentclass[reprint,12pt]{elsarticle}
\usepackage{amsmath}

\journal{Chaos, Solitons \& Fractals}
\newcommand{\sech}{\mbox{sech}}

\newcommand{\del}{\delta}
\newcommand{\al}{\alpha}

\newcommand{\lam}{\lambda}

\newcommand{\ga}{\gamma}

\newcommand{\bea}{\begin{eqnarray}}
	\newcommand{\eea}{\end{eqnarray}}
\newcommand{\bes}{\begin{subequations}}
	\newcommand{\ees}{\end{subequations}}

\usepackage{xcolor}

\begin{document}
	
	\begin{frontmatter}
		
		\title{Scalar and vector electromagnetic solitary waves in nonlinear hyperbolic media}
		
		\author[ku]{M. Kirane}
		\ead{mokhtar.kirane@ku.ac.ae; mkirane@univ-lr.fr}
		\author[ku]{S. Stalin\corref{auth2}}
		\cortext[auth2]{Corresponding author}
		\ead{stalin.seenimuthu@ku.ac.ae; stalin.cnld@gmail.com}
		\address[ku]{Department of Mathematics, College of Art and Sciences, Khalifa University of Science and Technology, Abu Dhabi, 127788, United Arab Emirates}
		
		
		
		
		\begin{abstract}
			In this paper, we investigate the problem of electromagnetic wave propagation in hyperbolic nonlinear media. To address this problem, we consider the scalar hyperbolic nonlinear Schr\"odinger system and its coupled version, namely hyperbolic Manakov type equations. These hyperbolic systems are shown to be non-integrable. Then, we examine the propagation properties of both the scalar and vector electromagnetic solitary waves by deriving their exact analytical forms through the Hirota bilinear method. A detailed analysis shows that the presence of hyperbolic transverse dispersion provides an additional degree of freedom to prevent the formation of singularity in both the scalar and vector solitary wave structures in this hyperbolic nonlinear media. Besides this, we realize that the solitary waves in this media possess fascinating propagation properties which cannot be observed in conventional nonlinear media. We believe that the present study will be very useful in analyzing electromagnetic wave propagation in hyperbolic nonlinear metamaterials.       
		\end{abstract}
		
		
	\end{frontmatter}
	
	\section{Introduction}\label{sec1}
	Recently, engineering the propagation of light using artifical material is one of the hot areas of research in nonlinear optics \cite{kivshar-hm}. In particular, metamaterials or left-handed materials, having double-negative refractive indices \cite{vaselago,sarsen}, are highly beneficial to construct  the meta-devices with desirable properties. An interesting aspect of this material is that one can tune its properties depending on a particular need or application. Such materials exhibit properties that are never or rarely observed in nature. To enhance these properties further, in recent years, highly anisotropic media with hyperbolic dispersion, which is characterized by the effective electric and magnetic tensors, have been introduced to the metamaterials \cite{kivshar-hm}. As a result, the resultant composite materials exhibit unique properties including enhanced superlensing effects \cite{kivshar-hm}, strong enhancement of spontaneous emission \cite{kivshar-hm}, and a large Purcell factor \cite{poddubny-1,poddubny-2,jacob}. These synthetic materials have also applications in sub-wavelength optics \cite{kivshar-hm}, heat transport \cite{biehs}, acoustics \cite{li}, and cosmology \cite{narimanov,cosmology}. The latter fascinating properties emerge from by engineering the following dispersion relation of extraordinary transverse magnetic polarized waves. It is represented by 
	\begin{equation}
		\frac{k_x^2+k_y^2}{\varepsilon_o}+\frac{k_z^2}{\varepsilon_e}=\bigg(\frac{\omega}{c}\bigg)^2,\label{1}
	\end{equation}         
	where, $k_x$, $k_y$, and $k_z$ are the $x$, $y$ and $z$ components of the wave vector $k$, respectively, $\omega$ is the frequency of the wave, $c$ is the speed of light,  
	$\varepsilon_o$, and $\varepsilon_e$ are the principal values of the permittivity of tensors. 
In the conventional dielectric medium, both of the latter quantities take positive values whereas in hyperbolic metamaterials they take opposite values. For instance, they can be either $\varepsilon_o>0$ and $\varepsilon_e<0$ or  $\varepsilon_o<0$ and $\varepsilon_e>0$. It is easy to produce the structures, having such principal values of permittivity tensors, that were first experimentally illustrated in magnetized plasma \cite{fisher}. Then, layered hyperbolic metamaterials have been created to demonstrate negative refraction and hyperbolic dispersion \cite{hoffman, leung, alu, vala}. These layered hyperbolic materials have been extensively studied by theoretically to explain the wave dispersion, refraction and propagation in such materials \cite{shen,orlov,chebykin}. It is noted that hyperbolic metamaterials are also created in the form of metallic nanowire arrays \cite{simovski}. It is interesting to point out that waveguides made from the hyperbolic media can act as non-magnetic left-handed media \cite{podolskiy}. Very recently, it has been predicted that a periodic arrangement of graphene layers behave as a hyperbolic metamaterials by tuning them from elliptic to hyperbolic dispersion through external voltage \cite{iorsh}. This is because graphene based hyperbolic metamaterial exhibits a strong Purcel effect.  In addition to these several other factors, the presence of hyperbolic dispersion and intrinsic nonlinearity makes the hyperbolic metamaterials as very interesting to investigate the problem of electromagnetic wave propagation in such a bulk nonlinear media \cite{ishii-1,ishii-2,assanto,maimistov-1,maimistov-2,maimistov-3}. 

To describe the electromagnetic wave propagation in an anisotropic nonlinear hyperbolic media, the following ($2+1$)-dimensional coupled hyperbolic nonlinear Schr\"odinger (CHNLS) equations have been derived from the Maxwell's equations under slowly varying envelope approximation \cite{maimistov}. They are written as follows:    
\begin{equation}
	i\frac{\partial q_j}{\partial z}+\frac{1}{2}\bigg(\frac{\partial^2 q_j}{\partial x^2}+\beta\frac{\partial^2 q_j}{\partial y^2}\bigg)+\gamma_j(\lvert q_1\rvert^2+\sigma \lvert q_2\rvert^2)q_j=0,~q_j\equiv q_j(x,y,z), ~j=1,2. \label{chnls}
\end{equation} 
Here, $q_j$'s are complex electric field envelopes of a transverse magnetic wave, $z$ is the normalized propagation distance, $x$ and $y$ are the transverse  coordinates. In water wave theory, $z$ represents the physical time, $x$ and $y$ denote the spatial coordinates. The dispersion coefficient $\beta$ is equal to $-1$  for a hyperbolic medium whereas it is equal to one for a conventional nonlinear medium.  The nonlinearity coefficients $\gamma_1=\frac{4\pi\chi^{(3)}}{\varepsilon_e}$,  and  $\gamma_2=\frac{4\pi\chi^{(3)}}{\varepsilon_o}$ along with $\sigma$ determine the self-focusing and self-defocusing nature of both the ordinary and hyperbolic nonlinear media. This can be decided by the signs of $\gamma_j$'s, which essentially depend on the values of $\varepsilon_e$ and $\varepsilon_o$. As we pointed out earlier, for the conventional media, $\varepsilon_e$ and $\varepsilon_o$ have the same sign, that is $\varepsilon_{e,o}>0$ or $\varepsilon_{e,o}<0$. However, for a hyperbolic medium, the signs of  $\varepsilon_e$ and $\varepsilon_o$ are opposite.  The latter  two choices imply that the CHNLS equations or hyperbolic Manakov type system of equations (\ref{chnls}) cover the problem of electromagnetic wave propagation in both the hyperbolic media and conventional anisotropic media. This consideration is very much essential in order to compare the solitary wave properties one from the other. However, we point out that the problem of dynamics of electromagnetic solitary waves in the hyperbolic media and their underlying mathematical structures have been focused to a lesser extent \cite{busch,baleanu-1,baleanu-2}. Therefore, the main objective of this paper is to unveil the electromagnetic solitary wave structures associated with the CHNLS Eq. (\ref{chnls}) by deriving their analytical forms through the Hirota's bilinear method and also analyze their propagation properties. We note that if the fields $q_1$ and $q_2$ do not vary with respect to $y$ or $x$ (or when $\beta=0$), Eq. (\ref{chnls}) becomes the well known Manakov type equations \cite{manakov} describing the evolution of orthogonally polarized light waves in  an ordinary anisotropic media or birefringent media. We also note that the real constant $\sigma=\pm 1$ help us to bring out the mixed focusing and defocusing Kerr nonlinearities in a hyperbolic medium in analogy with the mixed coupled NLS system. Due to this, the system (\ref{chnls}) with $\sigma=-1$ and $\gamma_j>0$, $j=1,2$, or with $\sigma=+1$ and  $\gamma_j<0$, $j=1,2$, are referred as the mixed CHNLS system. It is interesting to point out that several soliton structures are identified in the nonlinear Schr\"odinger family of systems \cite{28a,28b,28c,28d}.

Further, we also analyze a physical situation, in which an electromagnetic field has only one component. In other words, if the field depends on only one independent variable then under this situation, the wave propagation is described by the following scalar hyperbolic nonlinear Schr\"odinger (SHNLS) equation \cite{ablowitz}, 
\begin{equation}
	i\frac{\partial q}{\partial z}+\frac{1}{2}\bigg(\frac{\partial^2 q}{\partial x^2}+\beta\frac{\partial^2 q}{\partial y^2}\bigg)+\gamma\lvert q\rvert^2q=0,~q\equiv q(x,y,z). \label{shnls}
\end{equation}
The above SHNLS equation can be considered as a special case of the ($2+1$)-dimensional CHNLS Eq. (\ref{chnls}).  In Eq. (\ref{shnls}), the cubic Kerr nonlinearity coefficient $\gamma$ is defined in physical parameter as  $\displaystyle{\gamma=\frac{4\pi\chi^{(3)}}{\varepsilon_e}}$, where $\varepsilon_e$ is a value of permittivity tensor. For completeness, in the present paper, we also analyze the existence of electromagnetic solitary wave structures in the underlying SHNLS Eq. (\ref{shnls}) and reveal their properties. We note that the existence of travelling wave solution for Eq. (\ref{shnls}) with $\beta=-1$ was proved in Ref. \cite{saut}. We also note that the dynamics of spatial soliton inside a conventional Kerr nonlinear medium \cite{kivshar}. It is also interesting to point out that Eq. (\ref{shnls}), with $\beta=-1$, is also appears in the dynamics of gravity waves in deep-water \cite{stability-6}. 

The rest of the paper is organized as follows. In Section 2, we derive the scalar bright and dark solitary wave solutions and discuss their propagation properties. Then, the different types of vector solitary wave solutions of the CHNLS system (\ref{chnls}), such as bright-bright, bright-dark and dark-dark solitary wave solutions, are bring out in Section 3 and explore their properties. Further, we also indicate the stability analysis of solitary wave solutions in Section 4. Finally, we summarize our results in Section 5. In the Appendices, we present the Painlev\'e singularity structure analysis of the SHNLS Eq. (\ref{shnls}) and its coupled version (\ref{chnls}) and show that they are non-integrable.
\section{Scalar bright and dark electromagnetic solitary waves}
We start with the SHNLS Eq. (\ref{shnls}) and its bright and dark solitary wave solutions. To derive these solutions, we bilinearize Eq. (\ref{shnls}) by adopting the well known Hirota's bilinear method \cite{hirota}. We achieve this, through the bilinear transformation, $q(x,y,z)=g(x,y,z)/f(x,y,z)$, where $g(x,y,z)$ is a complex function and $f(x,y,z)$ is a real function. The following bilinear forms of the (2+1)-dimensional SHNLS Eq. (\ref{shnls}) are obtained by introducing the latter transformation in it. The resultant bilinear forms are given by
\begin{eqnarray}
	(iD_z+\frac{1}{2}(D_x^2+\beta D_y^2)-\lambda)g \cdot f=0, ~~(D_x^2+\beta D_y^2-2\lambda)f\cdot f=2\gamma \lvert g\rvert^2. \label{2}
\end{eqnarray}
In the above, $D_z$, $D_x$ and $D_y$ are the standard Hirota's $D$-operators. According to Hirota \cite{hirota}, these bilinear operators are defined as follows:
\begin{eqnarray}
	D_z^mD_x^n (a\cdot b)=\bigg(\frac{\partial}{\partial z}-\frac{\partial}{\partial z'}\bigg)^m\bigg(\frac{\partial}{\partial x}-\frac{\partial}{\partial x'}\bigg)^n a(z,x)b(z',x')_{ \lvert z=z',~x=x'}.
\end{eqnarray}
The above Hirota's notation is very useful to write the bilinear equation in  a more compact form. It is noted that the non-integrability nature of the scalar system (\ref{shnls}) and its coupled version (\ref{chnls}) is established by us and the details are given in Appendix A and B, respectively. Therefore, in the following sections, we refer the wave solutions associated with these systems as solitary wave solutions rather than calling them as `soliton solutions'. 
\subsection{Scalar bright solitary wave }
We consider a vanishing boundary condition $q\rightarrow 0$ as $x$ and $y$ go to $\pm \infty$ in order to derive the bright solitary wave solution of Eq. (\ref{shnls}). To incorporate this condition in the solution construction process, we fix $\lambda=0$ in the bilinear Eq. (\ref{2}). Solving the resultant bilinear forms using the truncated series expansion of the unknown functions $g$ and $f$, 	$g=\epsilon g_1$ and $f=1+\epsilon^2f_2$, where $\epsilon$ is a formal series expansion parameter, we arrive at the fundamental bright solitary wave solution of Eq. (\ref{shnls}). It is written as
\begin{subequations}
	\begin{eqnarray}
		&&q(x,y,z)=\frac{\epsilon g_1}{1+\epsilon^2f_2}=\frac{\al_1e^{\eta_1}}{1+e^{\eta_1+\eta_1^*+\del}},\label{6a}\\
		&&\eta_1=k_1x+\beta k_2y+\frac{i}{2}(k_1^2+\beta^3 k_2^2)z,~e^{\del}=\frac{\gamma\lvert \al_1\rvert^2}{(k_1+k_1^*)^2+\beta^3(k_2+k_2^*)^2}.\label{6b}
	\end{eqnarray}
\end{subequations}
From the above solution, one can observe that the structure of bright solitary wave in a hyperbolic medium is determined by the complex constants $\alpha_1$ and $k_j$, $j=1,2$, and a real constant $\gamma$, apart from a hyperbolic transverse dispersion coefficient $\beta=-1$. We point out that the same number of complex and real constants describe the structure of bright solitary wave in a conventional isotropic nonlinear medium but with the positive transverse dispersion ($\beta=1$) in $y$-direction. We also wish to point out that by repeating the same solution construction procedure with appropriate forms of seed solutions one cannot obtain the two-and general higher-order solitary wave solutions of Eq. (\ref{shnls}) with a sufficient number of free parameters because the series expansions do not terminate to a finite order thereby ensuring the non-integrable nature of Eq. (\ref{shnls}). 

To bring out the propagation properties associated with the bright solitary wave solution (\ref{6a})-(\ref{6b}), we rewrite it in the following hyperbolic form:
\begin{equation}
	q(x,y,z)=Ae^{i(\eta_{1I}+\theta)}\sech(\eta_{1R}+\frac{\del}{2}).\label{7}
\end{equation} 
Here, $A=\bigg(\frac{k_{1R}^2+\beta^3 k_{2R}^2}{\gamma}\bigg)^{1/2}$,  $e^{i\theta}=\sqrt{\al_1/\al^*_1}$, $\eta_{1R}=k_{1R}x+\beta k_{2R}y-\big(k_{1R}k_{1I}+\beta^3 k_{2R}k_{2I}\big)z$, $\eta_{1I}=k_{1I}x+\beta k_{2I}y+\frac{1}{2}\big((k_{1R}^2-k_{1I}^2)+\beta^3(k_{2R}^2-k_{2I}^2)\big)z$. The  subscripts $R$ and $I$ appearing in the above and in the following   represent the real and imaginary parts of that particular complex parameter. The amplitude of the bright solitary wave is $\displaystyle{\big(\frac{k_{1R}^2+\beta^3 k_{2R}^2}{\gamma}\big)^{1/2}}$. For $\beta=0$, the solution (\ref{7}) matches with the scalar bright soliton solution of the standard NLS equation \cite{zakharov}.  In the scalar Eq. (\ref{shnls}), the so-called line or stripe-solitary wave can travel in three different planes, namely ($x-y$)-plane, ($x-z$)-plane, and ($y-z$)-plane, for fixed $z$, $y$, and $x$, respectively. In ($x-y$)-plane , the solitary wave travels with velocity $\beta k_{2R}/k_{1R}$ whereas it propagates in ($x-z$)-plane with velocity $\displaystyle{k_{1I}+\frac{\beta^3 k_{2R}k_{2I}}{k_{1R}}}$. On the other hand,  the solitary wave moves in ($y-z$)-plane with velocity $\displaystyle{\frac{ k_{1R}k_{1I}}{\beta k_{2R}}+\beta^2 k_{2I}}$. From the latter, one can observe that the velocity of solitary wave in both planes depends on the real and imaginary parts of wave numbers $k_1$ and $k_2$. This implies that one cannot tune the velocity of solitary wave in a particular plane (say ($x-z$)-plane) without affecting the velocity of solitary wave in the other plane (say ($y-z$)-plane). This fact is always true in both the conventional and hyperbolic nonlinear media. 

\begin{figure}[]
	\centering
	\includegraphics[width=1.0\linewidth]{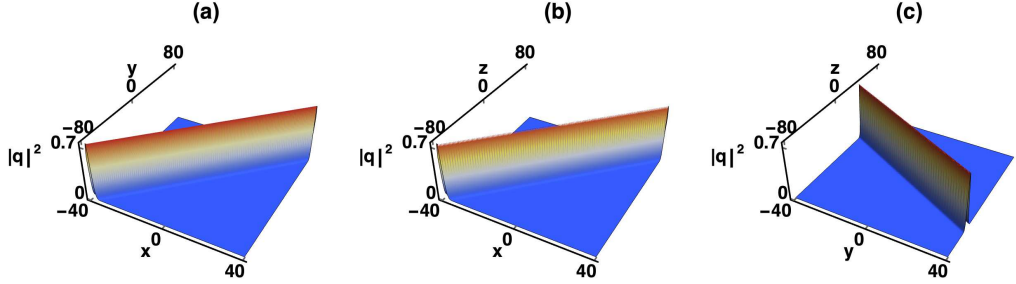}
	\caption{A regular scalar bright solitary wave propagation in three different planes is demonstrated for fixed $z$, $y$ and $x$, respectively. (a): ($x-y$) plane for $z=0$, (b): ($x-z$) plane for $y=0$, (c): ($y-z$) plane for $x=0$. The parameter values are $k_1=1+i$, $k_2=0.5+i$, $\alpha_1=1$, $\gamma=1$, $\beta=-1$.  }
	\label{fg1}
\end{figure}

Further, an interesting fact about the hyperbolic nonlinear media is that one can observe the bright solitary wave formation in focusing as well as defocusing Kerr nonlinear regimes. This fact can be realized by carefully looking at the solution (\ref{6a})-(\ref{6b}) of Eq. (\ref{shnls}) since it satisfies Eq. (\ref{shnls}) in both the focusing ($\gamma>0$) and defocusing ($\gamma<0$) nonlinear regions with $\beta=-1$. As a result, one can observe a non-singular bright solitary wave profile in such a hyperbolic medium by imposing either $k_{1R}^2>k_{2R}^2$,  $\gamma>0$ or  $k_{1R}^2<k_{2R}^2$,  $\gamma<0$. In both situations,  the quantity $e^{\delta}$ is always positive in Eq. (\ref{6b}).  However, the above fact is not true in the case of conventional  nonlinear medium, where a regular bright solitary wave exist only in the focusing nonlinear region. From the solution (\ref{6a})-(\ref{6b}), we realize  that it is always singular for $\beta=1$ and $\gamma<0$. It physically implies that the presence of hyperbolic dispersion in the nonlinear hyperbolic metamaterials provides an additional degree of freedom to avoid the formation of singularity in bright solitary wave structures in both the self-focusing and self-defocusing Kerr nonlinear regimes. Such a property is not possible in the conventional Kerr nonlinear medium. A typical propagation of non-singular bright solitary wave in a hyperbolic focusing nonlinear medium is illustrated in Figs. \ref{fg1}a-\ref{fg1}c. We have omitted a graphical demonstration of the propagation of the bright solitary wave in a hyperbolic defocusing Kerr nonlinear medium for brevity. 
\subsection{Scalar dark solitary wave}
\begin{figure}[]
	\centering
	\includegraphics[width=1.0\linewidth]{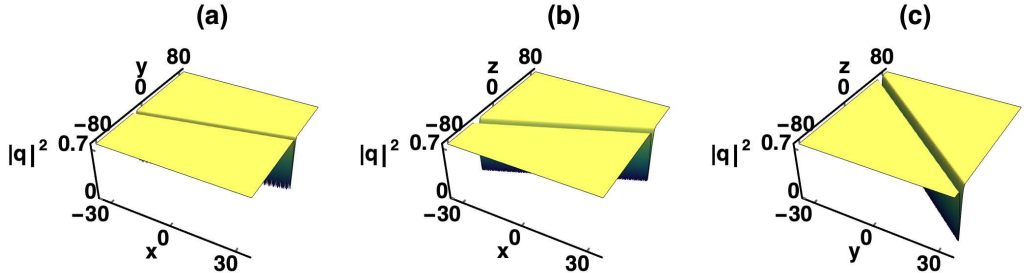}
	\caption{Propagation of scalar dark-solitary wave in the SHNLS equation. The parameter values are $k_1=1$, $k_2=2$, $l_1=1.3$, $l_2=0.2$, $\Omega_1=1$, $\gamma=-1$, $\beta=-1$. (a) ($x-y$)-plane: $z=0$, (b) ($x-z$)-plane: $y=0$, (c) ($y-z$)-plane: $x=0$.}
	\label{fg2}
\end{figure}

Next, to derive the dark-solitary wave solution of the system (\ref{shnls}): we wish to incorporate the non-vanishing boundary condition $q\rightarrow \tau e^{i\theta}$, $\theta=l_1x+\beta l_2y-\omega z$, as $x$ and $y$ go to $\pm\infty$ in the solution derivation procedure by setting $\lambda\neq 0$, which can be determined later, in the bilinear forms (\ref{2}). Then, solving Eq. (\ref{2}) along with the truncated series expansions, $g=g_0(1+\epsilon g_1)$, $f=1+\epsilon f_1
$, for the unknown functions, $g$ and $f$, we obtained the fundamental dark solitary wave solution of Eq. (\ref{shnls}) as
\begin{subequations}
	\begin{eqnarray}
		q(x,y,z)&=&\tau e^{i\theta}\frac{(1+Z_1e^{\xi_1})}{1+e^{\xi_1}},~~\xi_1=k_1x+\beta k_2y-\Omega_1z+\xi_1^{(0)},\label{8a}\\
		Z_1&=&-\frac{i(\Omega_1-l_1k_1-\beta^3 l_2k_2)+\frac{1}{2}(k_1^2+\beta^3 k_2^2)}{-i(\Omega_1-l_1k_1-\beta^3 l_2k_2)+\frac{1}{2}(k_1^2+\beta^3 k_2^2)},\label{8b}
	\end{eqnarray}
\end{subequations}
where $\theta=l_1x+\beta l_2y-\frac{1}{2}(l_1^2+\beta^3 l_2^2)z-\lam z$, $\lambda=-\gamma \tau^2$, along with a constraint equation for $\tau$:
\begin{equation}
	\tau=\pm\sqrt{\frac{(k_1^2+\beta^3 k_2^2)}{\gamma (-2+Z_1+Z_1^*)}}.
\end{equation}
In the above, the wave parameters, $\tau$, $k_j$, $l_j$, $j=1,2$, $\Omega_1$ and $\xi_1^{(0)}$ are real constants. The amplitude of the background plane wave $\tau$ can be a complex constant but in the present case we treat it as a real one.  
By rewriting the solution (\ref{8a}), we  obtain the following form of the fundamental dark solitary wave solution. It reads as 
\begin{eqnarray}
	q(x,y,z)=\frac{\tau}{2}e^{i\theta}\big[(1+Z_1)-(1-Z_1)\tanh\frac{\xi_1}{2}\big].\label{10}	
\end{eqnarray}
Here,  an interesting fact associated with dark-solitary wave solution (\ref{10}) is that it appears also in the focusing nonlinear region. This possibility is allowed in a hyperbolic nonlinear medium. The intensity of the scalar dark solitary wave can be calculated from Eq. (\ref{10}) as \cite{kivshar},
\begin{equation}
	\lvert q\rvert^2=\tau^2\bigg(1-X\sech^2\frac{\xi_1}{2}\bigg),\nonumber
\end{equation}
where
\begin{equation*}
	X=\frac{(k_1^2+\beta^3 k_2^2)^2}{\lvert(k_1^2+\beta^3 k_2^2)+2i(k_1l_1+k_2l_2\beta^3-\Omega_1)\rvert^2}.
\end{equation*}
The darkness of the solitary wave (\ref{10}) can be controlled by tuning the value of $X$ appropriately using the wave parameters involve in it. If $X=1$, we get a dark solitary wave profile whereas a gray solitary wave profile appears for $X<1$. In the present scalar hyperbolic medium as well as in the conventional nonlinear medium, the profile of dark-solitary wave is always regular. The propagation of a such non-singular dark solitary wave, for $\beta=-1$, in three different planes is illustrated in Fig. \ref{fg2}. For instance, the dark solitary wave propagation in ($x-y$) plane is demonstrated in Fig. \ref{fg2}a for $z=0$, where it moves with velocity $\displaystyle{\beta k_2/k_1}$. Then, the dark-solitary wave propagates in ($x-z$)-plane, with velocity $\Omega_1/k_1$, is depicted in Fig. \ref{fg2}b. A typical dark-solitary wave dynamics in ($y-z$)-plane is illustrated in Fig. \ref{fg2}c, where it propagates with velocity $\Omega_1/\beta k_2$. From the above, we observe that one can tune the velocity of dark solitary wave in ($x-z$)-plane through wave numbers without affecting the velocity in ($y-z$)-plane for fixed value of $\Omega_1$. The hyperbolic dispersion $\beta$ only influences  the grayness of the dark solitary wave. Very recently, the grayness of the dark-solion in the generalized long-wave short-wave resonance interaction system has been studied in \cite{stalin0}  by tuning the system parameter. It is noted that one can study the interaction properties of these solitary wave solutions numerically by considering our solutions as initial conditions \cite{alrazi}. We point out that the dark-soliton solution of the defocusing NLS equation can be recovered by imposing $\beta=0$ in Eq. (\ref{10}).       
\section{Vector solitary waves of the CHNLS system (\ref{chnls})}  
Now, we intend to study the propagation properties of the various vector electromagnetic solitary waves in a hyperbolic nonlinear medium in which the electric field has two components. To bring out these properties, we now consider a coupled version of the SHNLS Eq. (\ref{shnls}). An important feature of the CHNLS system (\ref{chnls})  is that it admits a rich structure of vector solitary wave solutions \cite{viji}, such as bright-bright, bright-dark or dark-bright, dark-dark type solitary wave solutions depending on the choice of boundary conditions. To construct these solutions, we again consider the bilinearization method. The CHNLS Eq. (\ref{chnls}) can be bilinearized through the transformation, $q_j(x,y,z)=g^{(j)}(x,y,z)/f(x,y,z)$, $j=1,2$, where $g^{(j)}(x,y,z)$'s are in general complex functions and $f(x,y,z)$ is a real function. By substituting the latter dependent variable transformation in Eq. (\ref{chnls}), we arrive at the following forms of bilinear equations. They are written as
\begin{equation}
	D_1g^{(j)} \cdot f=0,~~D_2f\cdot f=2\gamma(\lvert g^{(1)}\rvert^2+\sigma \lvert g^{(2)}\rvert^2), ~~j=1,2,\label{11} 
\end{equation}
where $D_1\equiv (iD_z+\frac{1}{2}(D_x^2+\beta D_y^2)-\lambda)$, and $D_2\equiv (D_x^2+\beta D_y^2-2\lambda)$. We note here that the bilinearization of the CHNLS system (\ref{chnls}) enforces us to fix the nonlinear coefficients $\gamma_1$ and $\gamma_2$ as $\gamma_1=\gamma_2=\gamma$. By solving the above bilinear forms (\ref{11}) along with different forms of series expansions for the unknown functions, $g^{(j)}$, $j=1,2$, and $f$, we are able to derive the various types of vector solitary wave solutions of the CHNLS system (\ref{chnls}) as given below. 
\subsection{Bright-bright solitary wave}
To obtain the fundamental vector bright solitary wave solution of the hyperbolic coupled NLS system (\ref{chnls}), we assume the truncated series expansions, $
g^{(j)}(x,y,z)=\epsilon g_1^{(j)}$, $j=1,2$, $f(x,y,z)=1+\epsilon^2f_2$, and $\lambda=0$. By solving the system of equations, which arise while considering the latter series expansions in Eq. (\ref{11}), we obtain the bright-bright vector solitary wave solution of Eq. (\ref{chnls}) as
\begin{eqnarray}
	&&\hspace{-0.8cm}q_j(x,y,z)=\frac{\epsilon g_1^{(j)}}{1+\epsilon^2f_2}=\frac{\al_1^{(j)}e^{\eta_1}}{1+e^{\eta_1+\eta_1^*+R_1}}, ~j=1,2,\label{12}\\
	&&\hspace{-0.8cm}\eta_1=k_1x+\beta k_2y+\frac{i}{2}(k_1^2+\beta^3 k_2^2)z, ~e^{R_1}=\frac{\gamma(\lvert\al_1^{(1)}\rvert^2+\sigma\lvert\al_1^{(2)}\rvert^2)}{(k_1+k_1^*)^2+\beta^3 (k_2+k_2^*)^2}.\nonumber
\end{eqnarray}
In the above solution, the parameters, $\alpha_1^{(j)}$, and  $k_j$, $j=1,2$, are complex and the system parameters, $\gamma$, $\beta$, and $\sigma$ are real constants. If the complex constant $\al_1^{(1)}=\al_1^{(2)}$, then the component $q_1=q_2$. Correspondingly, the solution (\ref{12}) becomes a solution of the SHNLS (\ref{shnls}) with a modified form of the nonlinearity. We note that to derive the solution (\ref{12}), the seed solutions, $g^{(j)}_1=\al_1^{(j)}e^{\eta_1}$, have been considered during the solution construction process. Besides this, we have also tried to construct the recently identified nondegenerate vector bright soliton/solitary wave solutions \cite{stalin1,stalin2,stalin3,stalin4} for Eq. (\ref{chnls}) by assuming two distinct starting solutions, $g^{(j)}_1=\al_1^{(j)}e^{\eta_j}$, $\eta_j=k_jx+p_jy-\omega_jz$, $j=1,2$. However, the series expansions for the unknown functions $g^{(j)}$, $j=1,2$, and $f$ do not converge to a finite order so we are unable to find such nondegenerate vector bright solitary wave solution for Eq. (\ref{chnls}) through the Hirota bilinear method.  We remark that one cannot find the higher-order solitary wave solutions of kind (\ref{12}), with a sufficient number of wave parameters, since the series expansions that we consider for the unknown functions $g^{(j)}$, $j=1,2$, and $f$ do not terminate to a finite order. This is due to the non-integrable nature of the CHNLS system (\ref{chnls}) as we proved it in Appendix B.       

\begin{figure}[]
	\centering
	\includegraphics[width=1.0\linewidth]{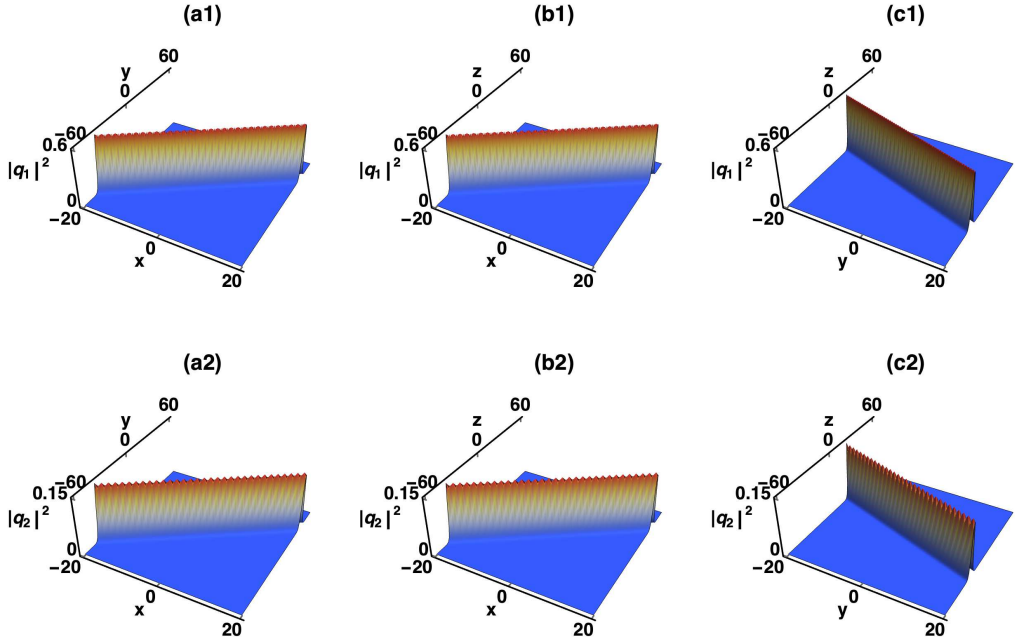}
	\caption{Propagation of vector bright solitary waves in the CHNLS equations. The parameter values are $k_1=1+i$, $k_2=0.5+i$, $\alpha_1^{(1)}=1$, $\alpha_1^{(2)}=0.5$, $\sigma=1$, $\gamma=1$, $\beta=-1$. (a1)-(a2) ($x-y$)-plane: $z=0$, (b1)-(b2) ($x-z$)-plane: $y=0$, (c1)-(c2) ($y-z$)-plane: $x=0$.}
	\label{fg3}
\end{figure}

In order to reveal the propagation properties associated with the vector bright solitary wave, we rewrite the solution (\ref{12}) as follows: 
\begin{eqnarray}
	q_j(x,y,z)=\hat{A}_j(\frac{k_{1R}^2+\beta^3 k_{2R}^2}{\gamma})^{1/2}e^{i\eta_{1I}}\sech(\eta_{1R}+\frac{R_1}{2}),\label{13}
\end{eqnarray}
where 
\begin{eqnarray*}
	&&\hat{A}_j=\frac{\al_1^{(j)}}{\sqrt{\lvert\al_1^{(1)}\rvert^2+\sigma\lvert\al_1^{(2)}\rvert^2}},~~~~ \frac{R_1}{2}=\frac{1}{2}\log\frac{\gamma(\lvert\al_1^{(1)}\rvert^2+\sigma\lvert\al_1^{(2)}\vert^2)}{(k_1+k_1^*)^2+\beta^3(k_2+k_2^*)^2},\\ &&\eta_{1R}=k_{1R}x+\beta k_{2R}y-\big(k_{1R}k_{1I}+\beta^3 k_{2R}k_{2I}\big)z,\\ &&\eta_{1I}=k_{1I}x+\beta k_{2I}y+\frac{1}{2}\big((k_{1R}^2-k_{1I}^2)+\beta^3(k_{2R}^2-k_{2I}^2)\big)z.
\end{eqnarray*}
The amplitudes of the bright solitary wave in both the components $q_1$ and $q_2$ are given by $\displaystyle{\hat{A}_1\big(\frac{k_{1R}^2+\beta^3 k_{2R}^2}{\gamma}\big)^{1/2}}$, and $\displaystyle{\hat{A}_2\big(\frac{k_{1R}^2+\beta^3 k_{2R}^2}{\gamma}\big)^{1/2}}$, respectively. Here, ${A}=\begin{pmatrix}
	\hat{A}_1 &\hat{A}_2
\end{pmatrix}^{T}$ is a unit polarization vector, which obeys $A^{\dagger}BA=1$, $B=\begin{pmatrix}
	1 &0\\ 0&\sigma
\end{pmatrix}$. The complex constants $\al_1^{(j)}$'s play a crucial role in fixing the intensity of the vector bright line-solitary waves. In the vector Eq. (\ref{chnls}), the line solitary wave can propagates in three different planes, like in the case of scalar Eq. (\ref{shnls}), namely ($x-y$)-plane, ($x-z$)-plane, and ($y-z$)-plane, for fixed $z$, $y$, and $x$, respectively. The velocity of vector solitary wave in ($x-z$)-plane is $\displaystyle{k_{1I}+\frac{\beta^3 k_{2R}k_{2I}}{k_{1R}}}$, while the bright solitary wave travels in ($y-z$)-plane with velocity $\displaystyle{\frac{ k_{1R}k_{1I}}{\beta k_{2R}}+\beta^2 k_{2I}}$. We displayed these propagation scenarios of vector bright solitary waves in Fig. \ref{fg3}. We note that the constant, $\beta$, which explicitly appears on the velocity of solitary wave.   

Further, the singularity nature of the vector bright solitary wave (\ref{13}) can be understood from the quantity $e^{R_1}$, which mainly depends on the value of $\al_1^{(j)}$, $k_{jR}$, $j=1,2$, $\beta$, $\gamma$, and $\sigma$. In the hyperbolic nonlinear media, a regular bright solitary wave profile appears in different nonlinear regions by fixing  the conditions as follows: (i) Focusing nonlinear case ($\gamma,\sigma>0$): $k_{1R}^2>k_{2R}^2$, (ii) Defocusing nonlinear case ($\gamma,\sigma<0$):  $k_{1R}^2<k_{2R}^2$, $\lvert\alpha_1^{(1)}\rvert>\lvert\alpha_1^{(2)}\rvert$, (iii) Mixed nonlinear case ($\gamma>0$, $\sigma=-1$ or $\gamma<0$, $\sigma=1$): $k_{1R}^2>k_{2R}^2$, $\lvert\alpha_1^{(1)}\rvert>\lvert\alpha_1^{(2)}\rvert$ (or $k_{1R}^2<k_{2R}^2$). However, this is not true in the case of conventional anisotropic nonlinear medium, where a non-singular vector bright solitary wave exists in the focusing nonlinear regime whereas a bright solitary wave becomes singular in the defocusing nonlinear regime. Then, it is possible to observe a regular bright solitary wave profile in the conventional medium with mixed nonlinearity. For instance, the quantity $e^{R_1}$ becomes positive definite for $\gamma>0$ and $\sigma=-1$ (or $\gamma<0$ and $\sigma=-1$) when $\lvert\alpha_1^{(1)}\rvert>\lvert\alpha_1^{(2)}\rvert$ (or $\lvert\alpha_1^{(1)}\rvert<\lvert\alpha_1^{(2)}\rvert$).  From the above, one can clearly confirm that by setting the wave numbers $k_1$ and $k_2$, which determine the propagation behaviors of vector bright solitary waves along $x$ and $y$ directions, in such a way one can avoid the formation of singularity in all the nonlinear regions of the hyperbolic media.    
\begin{figure}[]
	\centering
	\includegraphics[width=1.0\linewidth]{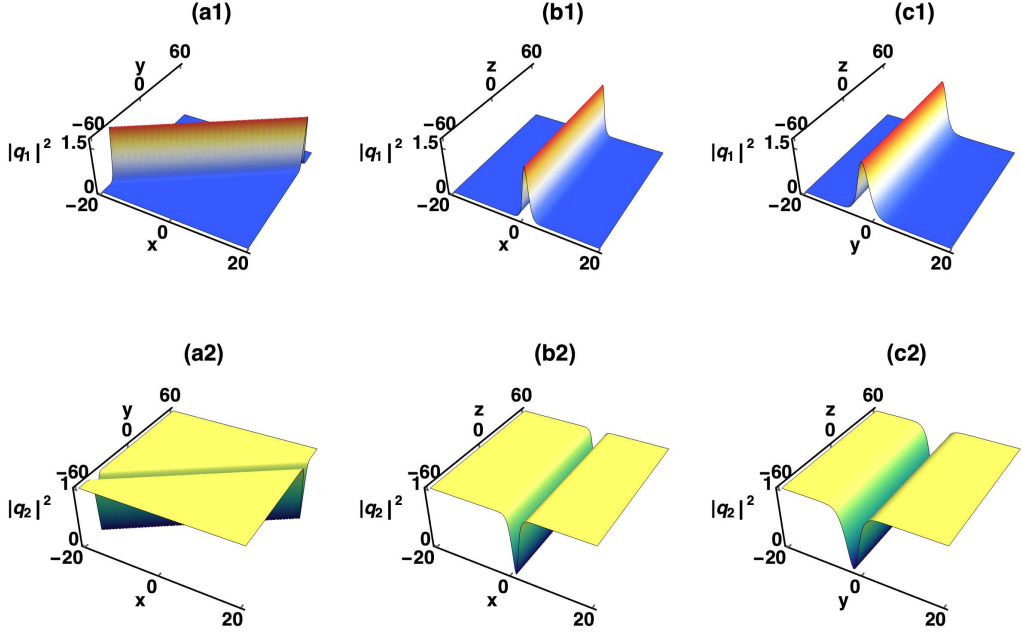}
	\caption{Propagation of bright-dark solitary waves in the CHNLS equations. The parameter values are $k_1=1+i$, $k_2=0.5+2i$, $\alpha_1^{(1)}=0.8$,  $\tau=1$, $\sigma=1$, $\gamma=1$, $\beta=-1$. (a1)-(a2) ($x-y$)-plane: $z=0$, (b1)-(b2) ($x-z$)-plane: $y=0$, (c1)-(c2) ($y-z$)-plane: $x=0$.}
	\label{fg4}
\end{figure}
\begin{figure}[]
	\centering
	\includegraphics[width=1.0\linewidth]{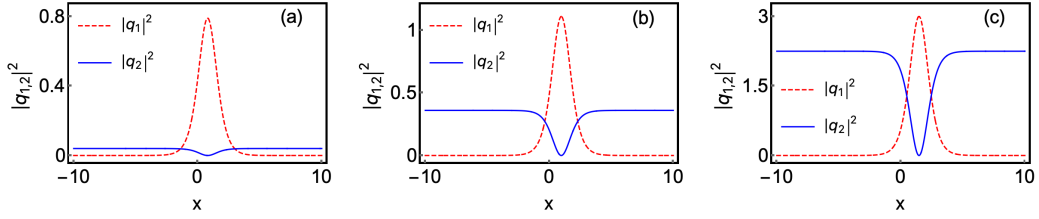}
	\caption{The role of dark-solitary wave component on the bright solitary wave component is displayed for various values of $\tau$ and for fixed values of other parameters. The parameter values are $k_1=1+i$, $k_2=0.5+2i$, $\alpha_1^{(1)}=0.8$,  $\sigma=1$, $\gamma=1$, $\beta=-1$. (a): $\tau=0.2$, (b): $\tau=0.6$, (c): $\tau=1.5$.}
	\label{fg5}
\end{figure}
\subsection{Bright-dark vector solitary wave}
To investigate the role of dark solitary wave on the bright solitary wave component, we intend to construct the fundamental mixed (bright-dark) vector  solitary wave solution of the system (\ref{chnls}). For this purpose,  we consider the series expansions as $ g^{(1)}=\epsilon g_1^{(1)}$, $g^{(2)}=g_0(1+\epsilon^2 g_2^{(2)})$, and  $f=1+\epsilon^2f_2$ along with the bilinear Eq. (\ref{11}) and $\lambda\neq 0$. In this case, the boundary conditions are assumed as $q_1\rightarrow 0$, and  $q_2\rightarrow \tau e^{i\gamma\sigma\tau^2z}$, as $x$ and $y$ go to $ \pm\infty$. By following the same solution construction procedure,  the fundamental bright-dark solitary wave solution is obtained as 
\begin{subequations}
	\begin{eqnarray}
		&&q_1=\frac{\epsilon g_1^{(1)}}{1+\epsilon^2f_2}=\frac{\al_1^{(1)}e^{\eta_1}}{1+e^{\eta_1+\eta_1^*+R_2}},\label{14a}\\
		&&q_2=g_0\frac{(1+\epsilon^2g_2^{(2)})}{1+\epsilon^2f_2}=\tau e^{i\ga\sigma\tau^2z}\frac{(1-e^{\eta_1+\eta_1^*+R_2+\chi_1})}{1+e^{\eta_1+\eta_1^*+R_2}},\label{14b}
	\end{eqnarray}
\end{subequations}
where 
\begin{eqnarray}
	&&\hspace{-0.5cm}\eta_1=k_1x+\beta k_2y+\frac{i}{2}(k_1^2+\beta^3 k_2^2)z-i\lambda z,~~\lambda=-\gamma\tau^2\sigma,\nonumber\\
	&&\hspace{-0.5cm}e^{\chi_1}=\frac{k_1(k_1+k_1^*)+\beta^3 k_2(k_2+k_2^*)}{k_1^*(k_1+k_1^*)+\beta^3 k_2^*(k_2+k_2^*)},~~e^{R_2}=\frac{\ga\lvert\al_1^{(1)}\rvert^2\chi_2}{(k_1+k_1^*)^2+\beta^3(k_2+k_2^*)^2},\nonumber\\
	&&\hspace{-0.5cm}\chi_2=\bigg(1+\frac{\sigma\tau^2\gamma\big((k_1+k_1^*)^2+\beta^3(k_2+k_2^*)^2\big)}{(k_1(k_1+k_1^*)+\beta^3 k_2(k_2+k_2^*))(k_1^*(k_1+k_1^*)+\beta^3 k_2^*(k_2+k_2^*))}\bigg)^{-1}.\nonumber
\end{eqnarray}
By rewriting the above solution, we arrive at the following compact form of the mixed vector solitary wave solution of Eq. (\ref{chnls}). It reads as
\begin{subequations}
	\begin{eqnarray}
		&&q_1(x,y,z)=A(\frac{k_{1R}^2+\beta^3 k_{2R}^2}{\gamma})^{1/2}e^{i(\eta_{1I}+\theta)}\sech(\eta_{1R}+\frac{R_2}{2}),\label{15a}\\
		&&q_2(x,y,z)=\tau e^{i(\ga\sigma\tau^2z+\phi+\pi)}\bigg[\cos\phi\tanh(\eta_{1R}+\frac{R_2}{2})+i\sin\phi\bigg],\label{15b}
	\end{eqnarray}
\end{subequations}
where 
\begin{eqnarray*}
	&&\eta_{1R}=k_{1R}x+\beta k_{2R}y-\big(k_{1R}k_{1I}+\beta^3 k_{2R}k_{2I}\big)z,\\ &&\eta_{1I}=k_{1I}x+\beta k_{2I}y+\frac{1}{2}\big[k_{1R}^2-k_{1I}^2+\beta^3(k_{2R}^2-k_{2I}^2)-\lambda\big]z, \\ &&\phi=\frac{1}{2}\tan^{-1}\bigg(\frac{2(k_{1R}^2+\beta^3 k_{2R}^2)(k_{1R}k_{1I}+\beta^3 k_{2R}k_{2I})}{(k_{1R}^2+\beta^3 k_{2R}^2)^2-(k_{1R}k_{1I}+\beta^3 k_{2R}k_{2I})^2}\bigg),~~e^{i\theta}=\sqrt{\frac{\al_1}{\al_1^*}},\\ &&A=\bigg[1+\frac{\sigma\tau^2\gamma\big((k_1+k_1^*)^2+\beta^3(k_2+k_2^*)^2\big)}{(k_1(k_1+k_1^*)+\beta^3 k_2(k_2+k_2^*))(k_1^*(k_1+k_1^*)+\beta^3 k_2^*(k_2+k_2^*))}\bigg]^{1/2}.
\end{eqnarray*}  The above mixed vector solution (\ref{15a})-(\ref{15b}) is characterized by three arbitrary complex constants, $\alpha_1$, $k_1$ and $k_2$, and a real constant $\tau$, apart from three system parameters, $\gamma$, $\sigma$, and $\beta$. The amplitude of the bright solitary wave is $\displaystyle{A(\frac{k_{1R}^2+\beta^3 k_{2R}^2}{\gamma})^{1/2}}$. Here, $A$ is related to the polarization vector of the bright solitary wave and is independent on the complex parameter $\alpha_1$, which appears as the complex phase $e^{i\theta}$. The bright and dark mixed vector solitary waves travel with velocity $\displaystyle{k_{1I}+\frac{\beta^3 k_{2R}k_{2I}}{k_{1R}}}$ in ($x-z$)-plane, while they propagate in ($y-z$)-plane with velocity $\displaystyle{\frac{ k_{1R}k_{1I}}{\beta k_{2R}}+\beta^2 k_{2I}}$. The propagation of bright and dark-solitary waves in three different planes is displayed in Fig. \ref{fg4}.     

\begin{figure}[]
	\centering
	\includegraphics[width=1.0\linewidth]{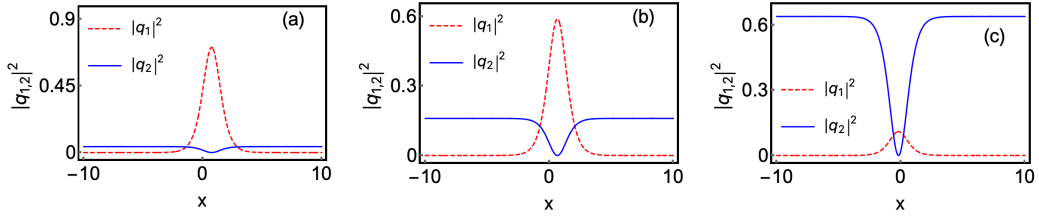}
	\caption{The role of dark-solitary wave on the bright solitary wave structure is illustrated. The parameter values are $k_1=1+i$, $k_2=0.5+2i$, $\alpha_1^{(1)}=0.8$,  $\sigma=-1$, $\gamma=1$, $\beta=-1$. (a) $\tau=0.2$, (b) $\tau=0.4$, (c) $\tau=0.8$.}
	\label{fg6}
\end{figure}

The intensity of dark-solitary wave component is calculated as $\lvert q_2\rvert^2=\tau^2\big(1-\cos^2\phi~\sech^2(\eta_{1R}+\frac{R_2}{2})\big)$. From the latter, one can observe that the value of $\phi$ determines the degree of grayness of the dark-solitary wave in $q_2$ component. Besides this, an interesting feature of the mixed vector solitary wave solution is that the presence of dark-solitary wave in $q_2$ component influences the structure of bright solitary wave in the other $q_1$ component. For instance, in the focusing and defocusing hyperbolic nonlinear media, the intensities of bright and dark solitary waves are increasing with respect to the background wave parameter $\tau$ as it is illustrated in Fig. \ref{fg5}. In addition to this, in the mixed nonlinear case, we observe that the intensity of the bright solitary wave is decreasing while increasing the value of $\tau$. Such a possibility is illustrated in Fig. \ref{fg6} for different values of $\tau$. From the solution (\ref{15a})-(\ref{15b}), it is obvious that the hyperbolic nonlinear media allows to observe the mixed bright-dark solitary waves.   
\begin{figure}[]
	\centering
	\includegraphics[width=1.0\linewidth]{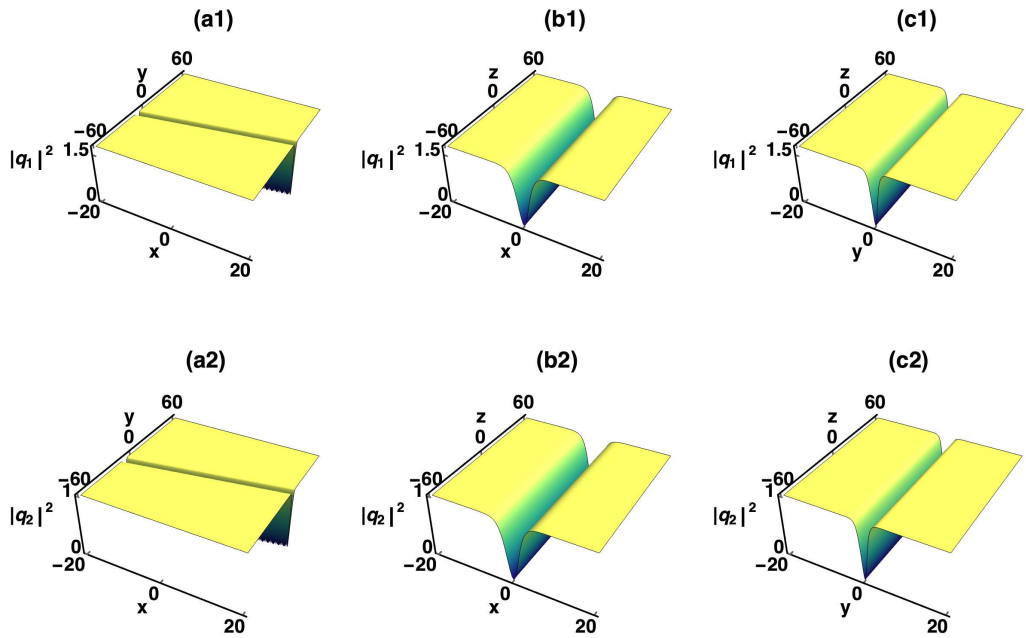}
	\caption{Propagation of vector dark solitary waves in the CHNLS equations. The wave parameter values are $k_1=1$, $k_2=2$, $l_1=1$, $l_2=0.3$, $l_3=0.7$, $l_4=0.2$, $\tau_2=1$, $\Omega_1=0.5$, $\sigma=1$, $\gamma=-1$, $\beta=-1$. (a1)-(a2) ($x-y$)-plane: $z=0$, (b1)-(b2) ($x-z$)-plane: $y=0$, (c1)-(c2) ($y-z$)-plane: $x=0$.}
	\label{fg7}
\end{figure}
\subsection{Dark-dark vector solitary wave}
Finally, to show the existence of the fundamental vector dark solitary wave in a hyperbolic medium, we derive the corresponding solution by considering the boundary condition: $\lvert q_j\rvert^2\rightarrow \tau_j^2$, $j=1,2$, as $x$ and $y$  go to $\pm \infty$. We derive this solitary wave solution by solving the bilinear Eq. (\ref{11}) along with the following truncated series expansions, $
g^{(1)}=g_0^{(1)}(1+\epsilon g_1^{(1)})$, $g^{(2)}=g_0^{(2)}(1+\epsilon g_1^{(2)})$, and $f=1+\epsilon f_1$. This action yields the fundamental vector dark-solitary wave solution of Eq. (\ref{chnls}). It reads as
\begin{eqnarray}
	&&q_j=\tau_je^{i\theta_j}\frac{(1+Z_je^{\xi_1})}{1+e^{\xi_1}},~~j=1,2,~~\xi_1=k_1x+\beta k_2y-\Omega_1z,\label{16}\\ 
	&&Z_1=-\frac{i(\Omega_1-l_1k_1-\beta^3 l_2k_2)+\frac{1}{2}(k_1^2+\beta^3 k_2^2)}{-i(\Omega_1-l_1k_1-\beta^3 l_2k_2)+\frac{1}{2}(k_1^2+\beta^3 k_2^2)},\nonumber\\
	&&Z_2=-\frac{i(\Omega_1-l_3k_1-\beta^3 l_4k_2)+\frac{1}{2}(k_1^2+\beta^3 k_2^2)}{-i(\Omega_1-l_3k_1-\beta^3 l_4k_2)+\frac{1}{2}(k_1^2+\beta^3 k_2^2)}.\nonumber
\end{eqnarray}
Here, $\theta_1=l_1x+\beta l_2y-\omega_1z$, $\theta_2=l_3x+\beta l_4y-\omega_2z$, $\omega_1=\frac{1}{2}(l_1^2+\beta^3 l_2^2)+\lambda$, $\omega_2=\frac{1}{2}(l_3^2+\beta^3 l_4^2)+\lambda$, $\lambda=-\gamma(\tau_1^2+\sigma\tau_2^2)$, 
and the constraint equation is given by
\begin{eqnarray}
	\tau_1=\pm\sqrt{\frac{\gamma\sigma\tau_2^2(-2+Z_2+Z_2^*)-(k_1^2+\beta^3 k_2^2)}{\gamma (2-Z_1-Z_1^*)}}.\nonumber
\end{eqnarray}
We rewrite the above dark-dark solitary wave solution (\ref{16}) as
\begin{subequations}
	\begin{eqnarray}
		q_1=\frac{\tau_1}{2}e^{i\theta_1}\bigg((1+Z_1)-(1-Z_1)\tanh\frac{\xi_1}{2}\bigg),\label{17a}\\
		q_2=\frac{\tau_2}{2}e^{i\theta_2}\bigg((1+Z_2)-(1-Z_2)\tanh\frac{\xi_1}{2}\bigg).\label{17b}
	\end{eqnarray}
\end{subequations}
The above vector dark solitary wave is characterized by 
the real parameters $k_j$, $l_j$, $\tau_j$, $j=1,2$, $l_3$, $l_4$, $\gamma$, $\beta$, and $\sigma$. The grayness of the dark-solitary wave in two components are determined by calculating intensity from the expressions (\ref{17a}) and (\ref{17b}). By doing so, we find
\begin{eqnarray}
	\lvert q_1\rvert^2=\tau_1^2 \big(1-X_1\sech^2\frac{\xi_1}{2}\big),~~	\lvert q_2\rvert^2=\tau_2^2 \big(1-X_2\sech^2\frac{\xi_1}{2}\big),
\end{eqnarray}
where 
\begin{subequations}
	\begin{eqnarray}
		X_1=\frac{(k_1^2+\beta^3 k_2^2)^2}{\lvert(k_1^2+\beta^3 k_2^2)+2i(k_1l_1+k_2l_2\beta^3-\Omega_1)\rvert^2},\\
		X_2=\frac{(k_1^2+\beta^3 k_2^2)^2}{\lvert(k_1^2+\beta^3 k_2^2)+2i(k_1l_3+k_2l_4\beta^3-\Omega_1)\rvert^2}.	
	\end{eqnarray}
\end{subequations}
One can get a complete dark-solitary wave profile in both components, if the above quantities $X_1$ and $X_2$ are equal to one. On the other hand, we get a gray-solitary wave profiles if $X_{1,2}<1$.  The various propagation of vector dark solitary wave, for $\beta=-1$, is illustrated in Fig. (\ref{fg7}). For example, the propagation of vector dark solitary wave in ($x-y$)-plane is demonstrated in Fig. (\ref{fg7})a1-a2, where it propagates with velocity $\displaystyle{\beta k_2/k_1}$. Then, the dark-solitary wave propagate in ($x-z$) plane for $y=0$, with velocity $\Omega_1/k_1$, is depicted in Fig. (\ref{fg7})b1-b2. A typical dark-solitary wave propagation in ($y-z$)-plane is displayed in Fig. (\ref{fg7})c1-c2, where it travels with velocity $\Omega_1/\beta k_2$. One can control the velocity of dark solitary wave, in both modes, in ($x-z$)-plane by tuning the values of wave numbers $k_1$ and $k_2$ without affecting the velocity in ($y-z$)-plane for fixed $\Omega_1$. The hyperbolic dispersion $\beta=-1$ is very useful to control the grayness of the vector dark solitary waves. We note that the vector dark solitary wave solution (\ref{16}) becomes a solution of the scalar Eq. (\ref{shnls}) if the restriction, $l_3=l_1$ and $l_4=l_2$, is imposed on it. Under this condition, $Z_2=Z_1$, the solution (\ref{16}) turns out to be $q_j=\frac{\tau_j}{2}e^{i\theta}\bigg(1+Z_1-(1-Z_1)\tanh\frac{\xi_1}{2}\bigg)$, where $\theta_1=\theta_2=\theta=l_1x+l_2y-\omega_1z$, $\omega_1=\frac{1}{2}(l_1^2+\beta l_2^2)+\lambda$, and $\lambda=-\gamma(\tau_1^2+\sigma\tau_2^2)$. 
\begin{figure}[]
	\centering
	\includegraphics[width=1.0\linewidth]{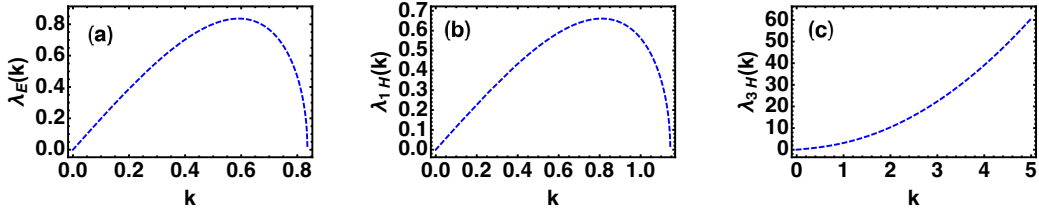}
	\caption{Fig. (a) demonstrates the growth rate $\lambda(k)$ for the scalar elliptic NLS case ($\beta=1$)  whereas Figs. (b) and (c) illustrate the growth rates of $\lambda_{1H}$ (real positive unstable eigenvalue) and imaginary part of $\lambda_{3H}$, respectively, for the scalar hyperbolic NLS case ($\beta=-1$).  }
	\label{fg8}
\end{figure}

\section{Transverse instability analysis}
We examine the stability of obtained scalar solitary wave solutions by treating the transverse dispersion term $\beta q_{yy}$  as perturbation to the $1$-dimensional scalar NLS soliton. To start with, we consider a very special case of the solution (\ref{7}) with $\beta=0$ and $z=0$ ($1$-dimensional stationary scalar NLS soliton). In other words, in order to examine the stability of 
\begin{equation}
q(x,y,z)=q_0(x)~e^{iz},\label{20}
\end{equation}
which satisfies 
\begin{equation}
q_{0,xx}-q_0+2q_0^3=0, \label{21}
\end{equation}
with respect to the transverse perturbations, we consider
\begin{equation}
q(x,y,z)=e^{i z}\bigg[q_0(x)+(v(x)+w(x))e^{\lambda z+iky}+(v^*(x)-w^*(x))e^{\lambda^* z-iky}\bigg].\label{22}
\end{equation}
In the above, $\lambda$ is a complex constant which is related to the spectral problem (Eq. (\ref{23})) given below, $k$ is the transverse wavenumber (which is real), and $v(x)$ and $w(x)<<1$ are small transverse perturbations. Here, we note that Eq. (\ref{21}) is obtained by substituting Eq. (\ref{20}), after taking $z\rightarrow 2z'$ and choosing $\gamma=1$, in Eq. (\ref{shnls}). Then, substituting (\ref{22}) in (\ref{shnls}), with again $z=2z'$ and $\gamma=1$, and linearize the resulting equation for $v(x)$ and $w(x)$, we get the following eigenvalue problem:
\begin{eqnarray}
\mathcal{L}_0w=-i\lambda v, ~~\mathcal{L}_1v=-i\lambda w, \label{23}
\end{eqnarray}  
where $\mathcal{L}_0=\partial_x^2-1+2 q_0^2-\beta k^2$, $\mathcal{L}_1=\partial_x^2-1+6 q_0^2-\beta k^2$, and $q_0=\sech x$. The above eigenvalue problem describes the stability of $1$-dimensional NLS soliton. 
One cannot find the exact analytical solutions for the linear eigenvalue problem (\ref{23}). However, this eigenvalue problem was extensively analysed in the literature both by analytically and numerically and reported the approximate growth rate for unstable eigenvalue $\lambda(k)$ and a threshold condition for the onset of the transverse instability \cite{stability-1,stability-2,stability-3,stability-4,stability-5,
stability-6,stability-7,stability-8,stability-8a,stability-9,stability-10,
stability-11,stability-12,stability-13}.  One may refer the paper (Ref. \cite{stability-13}) by Deconinck et al. for a complete analysis on the transverse instability of NLS soliton and also the previously achieved important results along this line are summarized there. Therefore, in order to avoid the repetition, we provide here only the expressions for the growth rates $\lambda(k)$, which was first obtained by Zakharov-Rubenchik \cite{stability-1}, for the long ($k\rightarrow 0$) and short ($k\rightarrow k_c$, here $k_c$ refers a critical value) transverse waves. Based on these expressions and their corresponding plots, we conclude the stability of our obtained solitary wave solutions. To achieve the expressions for approximate growth rates the asymptotic expansion method was adopted. This process yields the growth rate for the elliptic NLS case ($\beta=1$) is given by 
\begin{equation}
\lambda_E^2(k)=4k^2-\frac{4}{3}(\frac{\pi^2}{3}+1)k^4+O(k^6).\label{24}
\end{equation}
The above growth rate corresponds to the long transverse waves. However, for the short transverse waves one may refer \cite{stability-13}, where a cut-off condition for the onset of instability is given. We draw Eq. (\ref{24}) by varying the transverse wavenumber $k$ in Fig. \ref{fg8}(a), where the growth rate reaches its maximum value for $k\approx 0.6$.   

For the  hyperbolic case ($\beta=-1$), the approximate expressions for the four branches of $\lambda(k)$ have been obtained in \cite{stability-1,stability-13} by considering and long transverse waves as perturbations. Out of which two of them are imaginary whereas the other two is a real negative and a real positive value of $\lambda$. They are given as follows:  
\begin{equation}
\lambda_{1,2H}^2(k)=\frac{4}{3}k^2-\frac{4}{9}(\frac{\pi^2}{3}-1)k^4+O(k^6),\label{25}
\end{equation}
and 
\begin{equation}
\lambda_{3,4H}^2(k)=-4k^2-\frac{4}{3}(\frac{\pi^2}{3}+1)k^4+O(k^6).\label{26}
\end{equation}
In the above, $\lambda_{3,4H}$ denotes a pair of imaginary eigenvalues, $\lambda_{1H}$ and $\lambda_{2H}$  refer to the unstable (positive) and the stable (negative) eigenvalues, respectively, and $H$ represents the hyperbolic case. We draw the growth  rate $\lambda(k)$ corresponds to the unstable eigenvalue in Fig. \ref{fg8}(b) for certain values of $k$. However, there is a cut-off in instability curve after $k=1$. One can understand the reason behind this cut-off from Ref. \cite{stability-13}. Similarly, we also draw the imaginary part of growth rate $\lambda_{3H}$ in Fig. \ref{fg8}(c) for certain values of $k$. From these figures (and also from their corresponding growth rates expressions (\ref{25}) and (\ref{26})), and from Eq. (\ref{22}), we infer that the scalar NLS soliton $q_0=\sech ~x$ is unstable if eigenvalue $\lambda$ is real positive ($\lambda_{1H}$). This is due to the presence of exponential factor $e^{\lambda z}$ in Eq. (\ref{22}). Because of the latter factor the amplification will occur in the corresponding perturbation profile. On the other hand, if $\lambda$ is complex the effect of instability is determined by the real part of $\lambda_{3,4H}$ leading to the amplification of perturbation along with oscillatory instability which is due to the imaginary part of $\lambda_{3,4H}$. If the eigenvalue is a real negative value, then the NLS soliton is stable against the transverse perturbation. In a similar way, one can do the stability analysis for the rest of our solitary wave solutions. Thus, we conclude that the scalar and coupled solitary waves presented in this paper are also prone to perturbations in transverse directions leading to the transverse instability. 

.


\section{Conclusion}
In this paper, we have considered the scalar and vector hyperbolic nonlinear Schr\"odinger equations to study the propagation of electromagnetic solitary waves in nonlinear hyperbolic metamaterials. These equations are shown to be non-integrable since they fail to pass the Painlev\'e test. Then, we have investigated the propagation properties of both scalar and vector electromagnetic solitary waves by deriving their exact analytical forms through the Hirota bilinear method. In the scalar case, we have obtained the bright and dark solitary waves solutions and revealed their interesting propagation properties. On the other hand, in the vector case, we have presented the various types of vector solitary wave solutions, such as bright-bright, bright-dark and dark-dark solitary wave solutions of the CHNLS system (\ref{chnls}). Out of which, in the mixed vector solitary wave solution, the existence of dark solitary wave in one component strongly influences the structure of bright solitary wave 
in another component. Another interesting main result of our study is that the presence of negative/hyperbolic transverse dispersion in hyperbolic nonlinear media prevents the formation of wave collapsing phenomenon by avoiding the singularity formation. We observe that this is essentially due to the presence of two distinct transverse wavenumbers in the obtained solitary wave solutions. This striking property makes our solitary wave solutions as special. Then, the hyperbolic transverse dispersion also allows to observe bright and dark solitary waves in both focusing and defocusing Kerr nonlinear regions. We believe that the results reported in this paper will be useful for studying the electromagnetic wave propagation in the bulk nonlinear hyperbolic media.    

\section*{Declaration of competing interest }
The authors declare that they have no conflict of interest. 
\section*{Acknowledgement}
The works of Mokhtar Kirane, and Stalin Seenimuthu, are supported by the Khalifa University of Science and Technology (KUST), Abu-Dhabi, UAE, under the Project Grant No. 8474000355. 

\appendix

\section{Painlev\'e singularity structure analysis of SHNLS system (\ref{shnls})}
In this section, we intend to analyze the integrability of the system (\ref{shnls})  through the WTC (Weiss, Tabor and Carnevale) algorithm \cite{tabor}. To investigate integrability of the scalar hyperbolic NLS system (\ref{shnls}), we rename the nonlinear Schr\"odinger field $q(x,y,z)$ its complex conjugate field $q^*(x,y,z)$ as $a(x,y,z)$ and $b(x,y,z)$, respectively. Then Eq. (\ref{shnls}) and its complex conjugate equation can be rewritten as follows:  
\begin{subequations}
	\begin{eqnarray}
		ia_z+\frac{1}{2}(a_{xx}+\beta a_{yy})+\gamma a^2b=0, \label{20a}\\
		-ib_z+\frac{1}{2}(b_{xx}+\beta b_{yy})+\gamma b^2a=0. \label{20b}
	\end{eqnarray}
\end{subequations}
The Painlev\'e singularity structure analysis for the above coupled equations can be carried out by expanding their solutions in a standard way in the neighborhood of a non-characteristic manifold $\phi(x,y,z)=0$, with $\phi_x,\phi_y,\phi_z\neq 0$. The Laurent series expansion for the solutions $a(x,y,z)$ and $b(x,y,z)$ are 
\begin{subequations}
\begin{eqnarray}
	a(x,y,z)=\sum_{j=0}^{\infty}a_j(x,y,z)\phi^{j+\nu_1}, ~a_0\neq 0,\label{21a}\\
	b(x,y,z)=\sum_{j=0}^{\infty}b_j(x,y,z)\phi^{j+\nu_2}, ~b_0\neq 0,\label{21b}
\end{eqnarray}
\end{subequations}
where $\nu_k$, $k=1,2$, are integers, which are to be determined later. In the Painlev\'e test, the first step is to analyze the leading order behaviour of the solutions of Eqs. (\ref{20a}) and (\ref{20b}). To do this, we restrict the series expansion (\ref{21a}) and (\ref{21b}) as $a=a_0\phi^{\nu_1}$ and $b=b_0\phi^{\nu_2}$ by assuming $j=0$. The values of $\nu_1$ and $\nu_2$ are identified by substituting the latter leading order solutions in Eqs. (\ref{20a}) and (\ref{20b}) and balancing the most negative dominant terms. By doing so, we find $\nu_k=-1$, $k=1,2$, and  the following condition
\begin{equation}
	a_0b_0=-\frac{1}{\gamma}(\phi_x^2+\beta \phi_y^2 ). \label{a22}
\end{equation} 
From the above equation, one can observe that a single equation is obtained for the two unknowns $a_0$ and $b_0$, so one is arbitrary. The next step is to obtain the resonance values ($j$), at which  the arbitrary functions appear in the Laurent series (\ref{21a})-(\ref{21b}). This can be determined by substituting Eqs. (\ref{21a})-(\ref{21b}) in Eqs. (\ref{20a}) and (\ref{20b}) and collect the coefficients of $\phi^{j-3}$, which gives
\begin{eqnarray}
	\begin{pmatrix}
		(-(j-1)(j-2)+4)\gamma a_0b_0 & 2\gamma a_0^2\\
		 2\gamma b_0^2 &		(-(j-1)(j-2)+4)\gamma a_0b_0
	\end{pmatrix}\begin{pmatrix}
	a_j\\b_j
	\end{pmatrix}=0. \nonumber
\end{eqnarray}
From the above equation, one can get the quartic  polynomial in $j$ and solving such a polynomial leads to the resonance values: $j=-1,0,3$ and $4$. Here, the resonance value $j=-1$ represents the arbitrariness of the singular manifold $\phi(x,y,z)$ and $j=0$ denotes the arbitrariness of either $a_0$ or $b_0$, which can be confirmed from  Eq. (\ref{a22}).  
 
Next, we check the existence of sufficient number of arbitrary functions at the resonance values $j=3$ and $j=4$ without introduction of any movable singular manifold. To verify this, we first collect the coefficients of $\phi^{-2}$, which corresponds to the resonance value $j=1$. By doing so, we get the following equations which are expressed in matrix form as
\begin{eqnarray}
	\begin{pmatrix}
		A_1 & B_1\\
		 A_2 &	A_1
	\end{pmatrix}\begin{pmatrix}
	a_1\\b_1
	\end{pmatrix}=\begin{pmatrix}
	c_1 \\c_2
	\end{pmatrix}, \label{a23}
\end{eqnarray}
where $A_1=2\gamma a_0 b_0$, $A_2=\gamma b_0^2$, $B_1= \gamma a_0^2$, $c_1=ia_0\phi_z+\frac{1}{2}\big(2a_{0,x}\phi_x+a_0\phi_{xx}\big)+\frac{\beta}{2}\big(2a_{0,y}\phi_y+a_0\phi_{yy}\big)$, and $c_2=-ib_0\phi_z+\frac{1}{2}\big(2b_{0,x}\phi_x+b_0\phi_{xx}\big)+\frac{\beta}{2}\big(2b_{0,y}\phi_y+b_0\phi_{yy}\big)$. By solving, 
 Eq. (\ref{a23}), we obtain the expressions for $a_1$ and $b_1$. which are given below:
 \begin{eqnarray}
 a_1=\frac{c_1A_1-c_2B_1}{A_1^2-A_2B_1}, ~~b_1=\frac{c_2A_1B_1-c_1A_2B_1}{B_1(A_1^2-A_2B_1)}.
 \end{eqnarray}
 From the above, we confirm that there is no arbitrary function at this order. Similarly, we also confirm no arbitrary function enters into the Laurent series (\ref{21a})-(\ref{21b}) at the resonance value $j=2$. It can be verified by collecting the coefficients of $\phi^{-1}$. This action gives the matrix equation
 \begin{eqnarray}
 \begin{pmatrix}
 A_1 & B_1\\
 A_2 & A_1
 \end{pmatrix}\begin{pmatrix}
 a_2\\b_2
 \end{pmatrix}=\begin{pmatrix}
 d_1\\d_2
 \end{pmatrix},
 \end{eqnarray}
  where $d_1=-ia_{0,z}-\frac{1}{2}a_{0,xx}-\frac{\beta}{2}a_{0,yy}-\gamma\big(a_1^2b_0+2a_0a_1b_1\big)$, $d_2=ib_{0,z}-\frac{1}{2}b_{0,xx}-\frac{\beta}{2}b_{0,yy}-\gamma\big(a_0b_1^2+2b_0a_1b_1\big)$, and other elements are already defined above. By solving the above equation we obtain
  \begin{eqnarray}
  a_2=\frac{d_1A_1-d_2B_1}{A_1^2-A_2B_1}, ~~b_2=\frac{d_2A_1B_1-d_1A_2B_1}{B_1(A_1^2-A_2B_1)}.
\end{eqnarray}   
This shows that there is no arbitrary function at this stage. However, we observe the breaking of arbitrariness at the resonance value $j=3$. To ensure this, collect the coefficients of $\phi^{0}$. This yields the following equations, which are defined in matrix form:
\begin{eqnarray}
\begin{pmatrix}
b_0 & a_0\\
b_0 & a_0
\end{pmatrix}\begin{pmatrix}
a_3\\b_3
\end{pmatrix}=\begin{pmatrix}
c_1'\\c_2'
\end{pmatrix}.
\end{eqnarray} 
Here, $c_1'=\frac{1}{\gamma a_0}\big(-ia_{1,z}-ia_2\phi_z-\frac{1}{2}(a_{1,xx}+2a_{2,x}\phi_x+a_2\phi_{xx})-\frac{\beta}{2}(a_{1,yy}+2a_{2,y}\phi_y+a_2\phi_{yy})-\gamma(2a_1a_2b_0+2a_0a_2b_1+a_1^2b_1+2a_0a_1b_2)\big)$, and  $c_2'=\frac{1}{\gamma b_0}\big(ib_{1,z}+ib_2\phi_z-\frac{1}{2}(b_{1,xx}+2b_{2,x}\phi_x+b_2\phi_{xx})-\frac{\beta}{2}(b_{1,yy}+2b_{2,y}\phi_y+b_2\phi_{yy})-\gamma(2b_1b_2a_0+2b_0b_2a_1+b_1^2a_1+2b_0b_1a_2)\big)$. From the above one can easily identify that the elements $c_1'$ and $c_2'$ are distinct and they become identical if and only if $\beta=0$, which corresponds to the standard NLS equation. This clearly suggest that the scalar hyperbolic NLS equation (\ref{shnls}) fails to satisfy the Painlev\'e test at this order.

Now, we collect the coefficients of $\phi^1$ to check the arbitrariness of the functions $a_4(x,y,z)$ and $b_4(x,y,z)$ at the resonance value $j=4$. By doing so, we obtain
\begin{eqnarray}
\begin{pmatrix}
b_0 & -a_0\\
b_0 & -a_0
\end{pmatrix}\begin{pmatrix}
a_4\\b_4
\end{pmatrix}=\begin{pmatrix}
d_1'\\d_2'
\end{pmatrix},
\end{eqnarray} 
where $d_1'=\frac{1}{a_0}\big(ia_{2,z}+2ia_3\phi_z+\frac{1}{2}(a_{2,xx}+4a_{3,x}\phi_x+2a_3\phi_{xx})+\frac{\beta}{2}(a_{2,yy}+4a_{3,y}\phi_y+2a_3\phi_{yy})+\gamma(2a_1a_3b_0+2a_0a_3b_1+2a_0a_1b_3+b_0a_2^2+2a_0a_2b_2+a_1^2b_2+2a_1a_2b_1)\big)$, and $d_2'=\frac{1}{b_0}\big(ib_{2,z}+2ib_3\phi_z-\frac{1}{2}(b_{2,xx}+4b_{3,x}\phi_x+2b_3\phi_{xx})-\frac{\beta}{2}(b_{2,yy}+4b_{3,y}\phi_y+2b_3\phi_{yy})-\gamma(2b_1b_3a_0+2b_0b_3a_1+2b_0a_3b_1+a_0b_2^2+2b_0a_2b_2+a_2b_1^2
+2a_1b_1b_2)\big)$. This shows that the elements $d_1'$ and $d_2'$ are non-identical except for $\beta=0$. Thus, we confirm that the SHNLS system (\ref{shnls}) fails to pass the Painlev\'e analysis due to the non-existence of sufficient number of arbitrary functions in the Laurent series. Therefore, the SHNLS equation (\ref{shnls}) is not integrable in the Painlev\'e sense. 

\section{Singularity structure analysis of CHNLS system (\ref{chnls})}
Next, we move on proving the non-integrable nature of the coupled hyperbolic nonlinear Schr\"odinger system (\ref{chnls}). However, we omit the most of the mathematical details related to this singularity analysis due to the cumbersome nature of equations. To begin with, we rewrite the coupled system (\ref{chnls}) in terms of four real functions $P(x,y,z)$, $Q(x,y,z)$, $R(x,y,z)$, and $S(x,y,z)$, which are defined by $q_1=P+iQ$, and $q_2=R+iS$. Consequently, we get the following four coupled equations: 
\begin{subequations}
\begin{eqnarray}
-Q_z+\frac{1}{2}(P_{xx}+\beta P_{yy})+\gamma_1\big(P^2+Q^2+\sigma(R^2+S^2)\big)P=0,\label{A.10a}\\
P_z+\frac{1}{2}(Q_{xx}+\beta Q_{yy})+\gamma_1\big(P^2+Q^2+\sigma(R^2+S^2)\big)Q=0,\label{A.10b}\\
-S_z+\frac{1}{2}(R_{xx}+\beta R_{yy})+\gamma_2\big(P^2+Q^2+\sigma(R^2+S^2)\big)R=0,\label{A.10c}\\
R_z+\frac{1}{2}(S_{xx}+\beta S_{yy})+\gamma_2\big(P^2+Q^2+\sigma(R^2+S^2)\big)S=0.\label{A.10d}
\end{eqnarray}
\end{subequations}
As we have described above for the SHNLS equation, we have to perform the singularity structure analysis for the above set of coupled equations. To start with, we consider the leading orders of the form
\begin{equation}
P=P_0\phi^{\nu_1},~Q=Q_0\phi^{\nu_2},~R=R_0\phi^{\nu_3},~S=S_0\phi^{\nu_4},
\end{equation}   
where $\nu_j$, $j=1,2,3,4$ are integers to be determined later. Substituting the above leading order forms in Eqs. (\ref{A.10a})-(\ref{A.10d}) and equating the most dominant terms we obtain the following choice:
\begin{subequations}
\begin{eqnarray}
\gamma_1\big(P_0^2+Q_0^2+\sigma(R_0^2+S_0^2)\big)=-(\phi_x^2+\beta\phi_y^2),\label{A.12a}\\
\gamma_2\big(P_0^2+Q_0^2+\sigma(R_0^2+S_0^2)\big)=-(\phi_x^2+\beta\phi_y^2),\label{A.12b}
\end{eqnarray} \end{subequations}
and $\nu_1=\nu_2=\nu_3=\nu_4=-1$. The remaining set of equations resulting from Eqs. (\ref{A.10a})-(\ref{A.10d}) are identical to the above set of equations. At this stage, one can say that out of four functions, $P_0$, $Q_0$, $R_0$, and $S_0$, two of them are arbitrary. However, the obtained resonance values, $j=-1,0,0,0,3,3,3,4$, show that $Q_0$ (or $S_0$) is arbitrary besides $P_0$ and $R_0$. Thus, the obtained Eqs. (\ref{A.12a}) and (\ref{A.12b}) reduce to a single equation, $\gamma\big(P_0^2+Q_0^2+\sigma(R_0^2+S_0^2)\big)=-(\phi_x^2+\beta\phi_y^2)$ along with a condition $\gamma_1=\gamma_2=\gamma$. Then, we found that the non-existence of sufficient number of arbitrary functions at the resonance values $j=3$ and $j=4$. This indicates that the CHNLS system (\ref{chnls}) is also non-integrable in the Painlev\'e sense.   
\end{document}